\documentclass[aps,twocolumn,superscriptaddress,showpacs]{revtex4}
\usepackage{amsmath,graphicx}
\newlength{\figwidth}
\newlength{\figlarge}
\begin{document}
\title{Mott transition in the Hubbard model 
on the anisotropic kagom\'e lattice}
\author{Yuta Furukawa}
\affiliation{%
Department of Physics, Kyoto University, 
Kyoto 606-8502, Japan}%
\author{Takuma Ohashi}
\affiliation{%
Department of Physics, Osaka University, 
Toyonaka, Osaka 560-0043, Japan}%
\author{Yohta Koyama}
\affiliation{%
Department of Physics, Kyoto University, 
Kyoto 606-8502, Japan}%
\author{Norio Kawakami}
\affiliation{%
Department of Physics, Kyoto University, 
Kyoto 606-8502, Japan}%
\date{\today}
\begin{abstract}
We investigate the Mott transition in the anisotropic kagom\'e lattice Hubbard model using the cellular dynamical mean field theory combined with continuous-time quantum Monte Carlo simulations. By calculating the double occupancy and the density of states, we determine the interaction strength of the first-order Mott transition and show that it becomes small as the anisotropy increases. We also calculate the spin correlation functions and the single-particle spectrum, and reveal that the quasiparticle and magnetic properties change dramatically around the Mott transition; the spin correlations are strongly enhanced and the quasiparticle bands are deformed. We conclude that such dramatic changes are due to the enhancement of anisotropy associated with the relaxation of frustration around the Mott transition.

\end{abstract}
\maketitle

Geometrically frustrated electron systems have provided hot topics in the field of strongly correlated electron systems, and have uncovered various new aspects of the Mott transition. The discovery of heavy fermion behavior in $\mathrm{LiV_2O_4}$ \cite{kondo97,jonsson07} with pyrochlore lattice structure has activated theoretical studies of electron correlations with geometrical frustration \cite{imai02,arita07,yoshioka08,hattori09}. More remarkably, recent experiments on the triangular lattice organic materials $\kappa$-(BEDT-TTF)$_2\mathrm{X}$ have revealed a novel spin liquid ground state in the Mott insulating phase \cite{lefebvre00,shimizu03,kagawa04,kashima01,yoshioka09}.

\begin{figure}[bt]
\begin{center}
\includegraphics[clip,width=\figwidth]{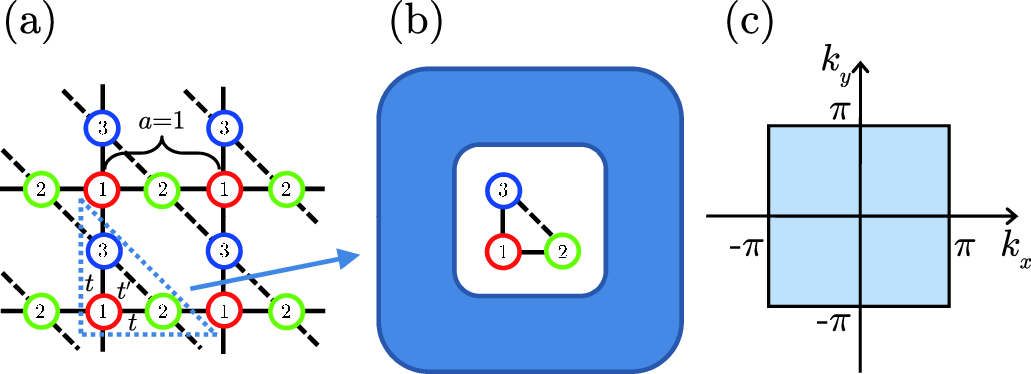}
\end{center}
\caption{(a) Anisotropic kagom\'e lattice, (b) effective cluster 
model used for the CDMFT and (c) first Brillouin zone of the 
anisotropic kagom\'e lattice (c). 
\label{fig:kagome}
}
\end{figure}

The kagom\'e lattice (see Fig. \ref{fig:kagome} (a)) is another prototype of frustrated systems, which shares some essential properties with other frustrated lattices. The localized electron systems on this lattice have been intensively studied and many unusual properties have been found \cite{misguich04}. 
In particular, a spin liquid state observed in the herbertsmithite $\mathrm{ZnCu3(OH)_6Cl_2}$ with kagom\'e lattice structure \cite{helton07, mendels07} has received considerable attention. A related spinel-oxide $\mathrm{Na_4Ir_3O_8}$ with hyperkagom\'e structure, which is a three-dimensional analog of kagom\'e lattice, is also proposed as a candidate for the spin liquid \cite{okamoto07,zhou08}. 
The issue of electron correlations on the kagom\'e lattice was addressed by applying the FLEX approximation \cite{imai03} and the QMC method \cite{bulut05} to the Hubbard model. Also in our recent study, we found a heavy Fermi liquid state emerging in a metallic phase near the Mott transition \cite{ohashi06}. Furthermore, an idea of chirality-spin separation was proposed to explain low-energy characteristic properties \cite{udagawa10,udagawa10p}. In spite of such intensive theoretical investigations, 
however, the effects of spatial anisotropy have not yet been addressed. The systematic exploration of the anisotropy is desired to elucidate how the frustration affects the nature of Mott transition, the quasiparticle formation, the magnetic properties, etc. It has indeed been clarified that in a moderately frustrated system on the anisotropic triangular lattice, the Mott transition shows a reentrant behavior \cite{kagawa04,ohashi08,liebsch09}, where the low-temperature transition is governed by the enhanced antiferromagnetic (AFM) correlations. This is a unique feature that may occur neither in the unfrustrated square lattice \cite{maier05} nor in the fully frustrated triangular lattice\cite{parcollet04}. This naturally motivates us to study the effects of frustration on the Mott transition by systematically controlling the spatial anisotropy. 

In this paper, we study the Mott transition in the Hubbard model on the anisotropic kagom\'e lattice by means of the cellular dynamical mean field theory (CDMFT) \cite{kotliar01} combined with the continuous-time quantum Monte Carlo (CT-QMC) method \cite{werner06}. We first treat the isotropic system to confirm our previous results studied with a slightly different method \cite{ohashi06}. As the anisotropy is introduced, electronic properties drastically change around the Mott transition; the spin correlations are strongly enhanced and the quasiparticle bands are deformed substantially even if the system is slightly away from the isotropic point.  We elucidate that such dramatic changes are due to the enhancement of anisotropy driven by the relaxation of frustration around the Mott transition.

We consider the Hubbard model on the anisotropic kagom\'e lattice shown in Fig.\ref{fig:kagome}(a). Here, we set the lattice geometry as a square form, where the first Brillouin zone is $-\pi<k_x, k_y<\pi$ as shown in Fig.\ref{fig:kagome}(c).  The Hamiltonian reads
\begin{align}
H=  -t\sum_{\left \langle i,j \right \rangle ,\sigma}
c_{i\sigma }^\dag c_{j\sigma}
-t'\sum_{( i,j ) ,\sigma}
c_{i\sigma }^\dag c_{j\sigma}
+ U \sum_{i} n_{i\uparrow} n_{i\downarrow} ,
\label{eqn:hm}
\end{align}
where $c_{i\sigma }^\dag$ ($c_{j\sigma}$) creates (annihilates) an electron with spin $\sigma$ at site $i$ and $n_{i\sigma}=c_{i\sigma}^\dag c_{i\sigma}$. The hopping integrals and the Hubbard interaction are denoted by $t$ $(t')$ and $U$, respectively. The system corresponds to the fully frustrated kagom\'e lattice at $t'/t = 1$, and frustration becomes weaker with decreasing $t'/t$. The end member at $t'/t=0$ is called a decorated square lattice. 
We choose the energy unit as $t=1$ hereafter. To investigate strong correlations and geometrical effects, we use CDMFT \cite{kotliar01}, a cluster extension of DMFT \cite{georges96}. CDMFT has been successfully applied to a lot of frustrated electron systems so far \cite{ohashi06,ohashi08,parcollet04,kyung06,Galanakis09}. 

In CDMFT, the original lattice problem is mapped onto an effective cluster problem. Each unit cell of the anisotropic kagom\'e lattice has three sites, as shown in Fig. \ref{fig:kagome}(a). We thus end up with a three-site cluster model coupled to a self-consistently determined medium illustrated in Fig. \ref{fig:kagome}(b). Given the Green's function for the effective medium, $\hat{\mathcal{G}}$, we compute the cluster Green's function $\hat{G}$ and the cluster self-energy $\hat{\Sigma}$ in the effective cluster model, where $\hat{O}$ denotes a $3 \times 3$ matrix.  
To calculate $\hat{G}$ and $\hat{\Sigma}$ in the effective cluster model, we use the CT-QMC method. 
In CT-QMC simulations for an effective cluster model, the sign problem strongly depends on the choice of basis set in the local Hamiltonian. We choose the basis set which diagonalizes the local hopping matrix $\hat{t}_\mathrm{loc}$ written in the sublattice basis, 
\begin{align}
\hat{t}_\mathrm{loc}=
\left(
\begin{array}{ccc}
0&t&t\\
t&0&t'\\
t&t'&0
\end{array}
\right). 
\end{align}
We find that this choice substantially reduces the sign problem in QMC simulations. We iterate the DMFT self-consistent loop until the convergence of this procedure is achieved within 50 iterations at most. In each iteration, we typically use $2.5\times 10^{7}$ QMC sweeps to reach sufficient computational accuracy at very low temperature, $T=0.05$.

\begin{figure}[bt]
\begin{center}
\includegraphics[clip,width=\figwidth]{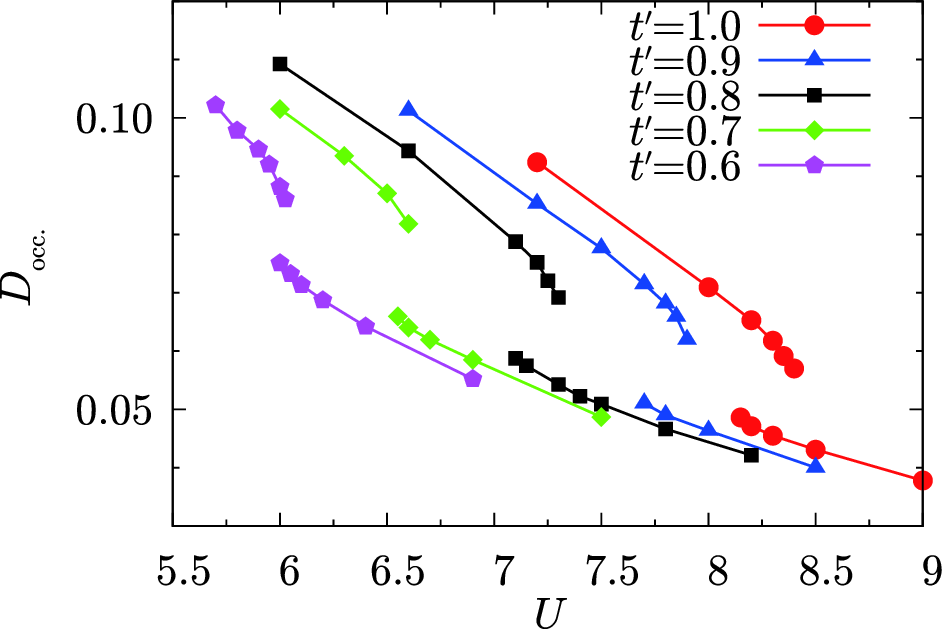}
\end{center}
\caption{
Double occupancy as a function of interaction strength $U$ 
for several $t'$ at $T=0.05$.}
\label{fig:double}
\end{figure}

Let us now investigate the Mott transition on the anisotropic kagom\'e lattice at half filling. 
In order to find evidence of the Mott transition, we calculate the $U$ dependence of the double occupancy $D_{occ.} = \frac{1}{3}\sum_{m=1}^3 
\left \langle n_{m\uparrow}n_{m\downarrow} \right \rangle$ 
for various $t'$. The double occupancy monotonically decreases with increasing $U$ and shows a discontinuous jump at the interaction $U_c$. We also find hysteresis, which signals the emergence of the first order Mott transition. In the isotropic kagom\'e system for $t'=1.0$, the Mott transition occurs at fairly large interaction strength $U_c \sim 8.4$ compared with the bandwidth $W=6$. This is consistent with our previous study using CDMFT combined with Hirsch-Fye QMC \cite{ohashi06}. In anisotropic cases, the interaction strength $U_c$ becomes small as $t'$ decreases. Note that the band width $W/t=4\sqrt{2} \sim 5.66$ for $t'=0$ and $W$ is almost unchanged when $t'$ varies. Therefore, we can say that the metallic region shrinks as geometrical frustration is weakened. 

\begin{figure}[bt]
\begin{center}
\includegraphics[clip,width=\figlarge]{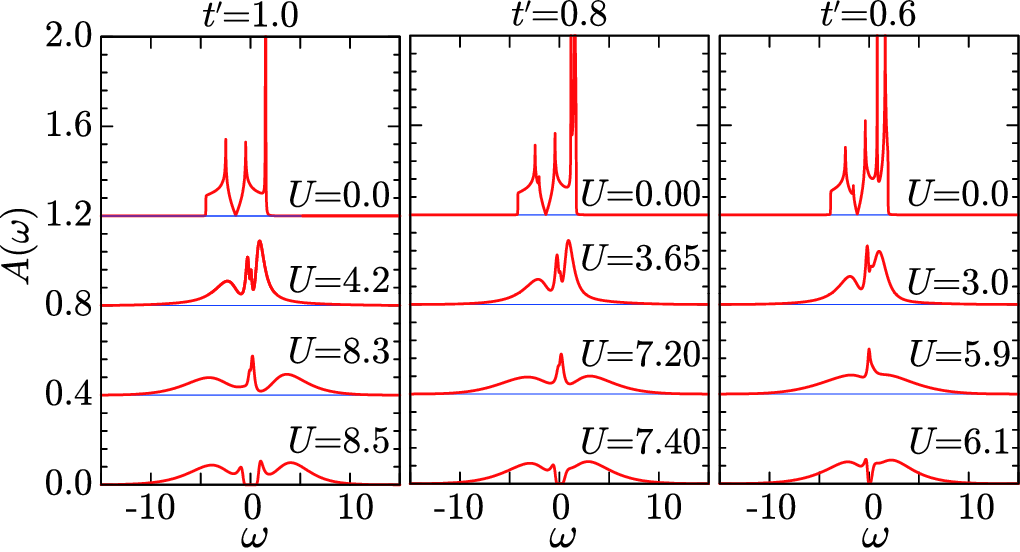}
\end{center}
\caption{
Density of states for typical $U$ and $t'$ at $T=0.05$. 
\label{fig:dos}}
\end{figure}

To see the metal-insulator transition more clearly, we calculate the density of states (DOS) $A(\omega)=-\frac{1}{3\pi}\sum_{m=1}^3\mathrm{Im}G_{mm}(\omega+i\delta) $ by applying the maximum entropy method (MEM) \cite{jarrell96} to the imaginary-time QMC data. In Fig.\ref{fig:dos}, we show the DOS for $t' = 1.0$, $0.8$ and $0.6$. We clearly see that the insulating gap opens around the Fermi level in the large $U$ region beyond $U_c$. For $t'=1.0$, the three bands are strongly renormalized and the heavy quasiparticle states emerge near the Fermi level in the strong correlation regime. A similar peak structure also appears for $t'=0.8, 0.6$  in the weaker $U$ region than that for $t'=1.0$. 

\begin{figure}[bt]
\begin{center}
\includegraphics[clip,width=\figwidth]{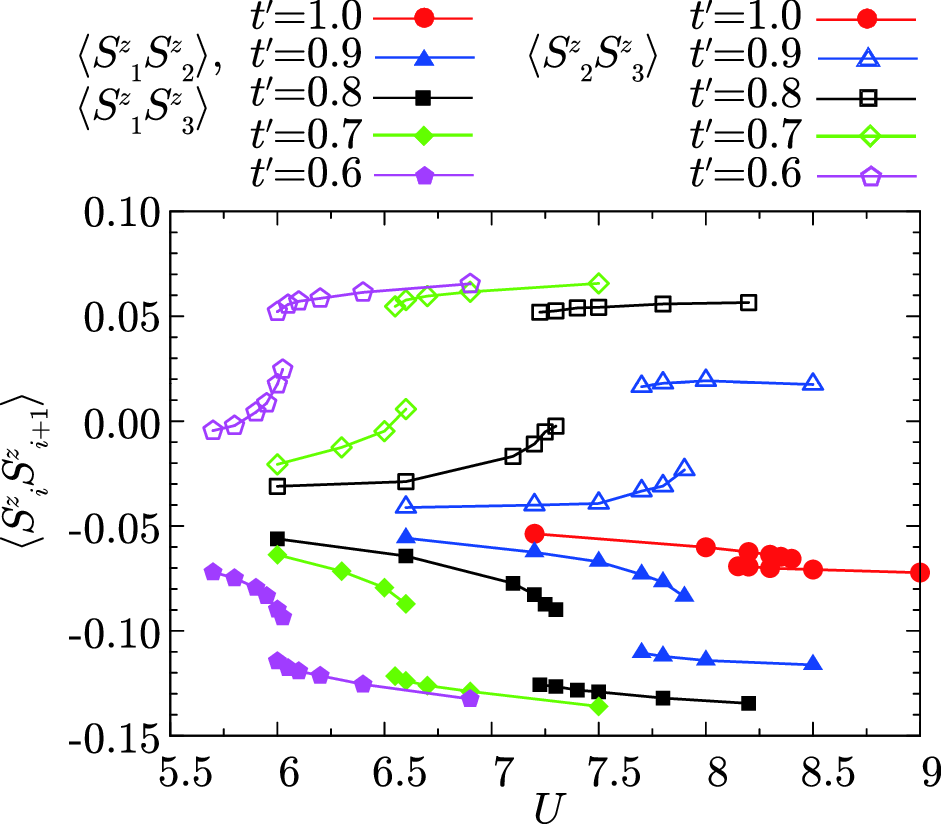}
\end{center}
\caption{
Nearest-neighbor spin correlation function 
$\langle S^z_i S^z_{i+1} \rangle$ as a function of $U$ for several $t'$, where the suffix $i$ specifies a sublattice as shown in Fig. \ref{fig:kagome}. 
\label{fig:scorr}}
\end{figure}

Let us now turn to the magnetic correlations. To see how frustration affects them around the Mott transition, we calculate the nearest-neighbor spin correlation function $\langle S^z_i S^z_{i+1}\rangle$.  Here, $i$ denotes a sublattice index as shown in Fig. \ref{fig:kagome}.  In Fig. \ref{fig:scorr}, we show 
$\langle S^z_i S^z_{i+1}\rangle$ 
as a function of $U$ for several $t'$. 
In the isotropic case, 
$\langle S^z_1 S^z_{2}\rangle$, 
$\langle S^z_1 S^z_{3}\rangle$ 
and $\langle S^z_2 S^z_{3}\rangle$ are equivalent and have small negative values for any $U$. This indicates that the AFM spin correlation is suppressed due to the strong frustration. As finite anisotropy is introduced ($t' = 0.9,0.8,0.7,0.6$), the spin correlations dramatically change near the Mott transition; 
$\langle S^z_1 S^z_{2}\rangle$ and $\langle S^z_1 S^z_{3}\rangle$ 
show strong AF correlations while 
$\langle S^z_2 S^z_{3}\rangle$ tends to be ferromagnetic. These behaviors indicate the tendency to a ferrimagnetic ordering (or an AFM-type ordering on a decorated square lattice). We thus find that magnetic properties in the system are very sensitive to the anisotropy, whose effect is particularly enhanced in the vicinity of the Mott transition point.

\begin{figure}[bt]
\begin{center}
\includegraphics[clip,width=\figlarge]{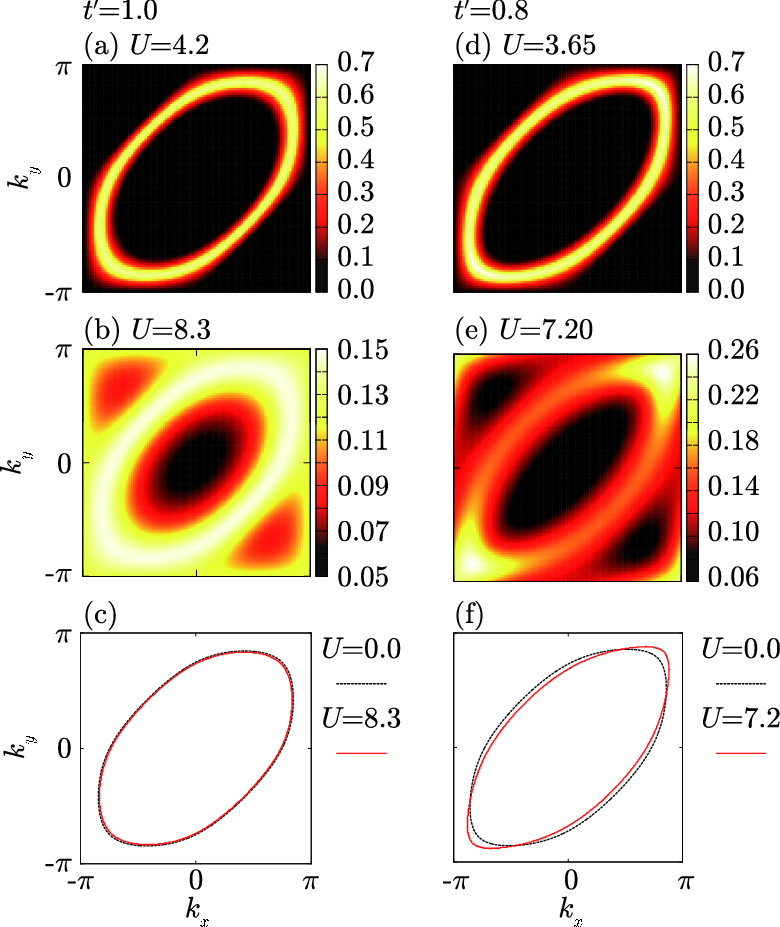}
\end{center}
\vspace{-6mm}
\caption{ Momentum resolved spectral weight at the Fermi level. In (c) and (f), we show the Fermi surfaces deduced from (b) and (e) together with those for the noninteracting case for reference.
\label{fig:Ak}
}
\end{figure}

The relaxation of frustration strongly affects the formation of quasiparticle bands. We next investigate the momentum resolved spectral weight at the Fermi level, $A(\mathbf{k},\omega=0)=-\frac{1}{3\pi}\sum_{m=1}^{3}
{\rm Im}G_{mm}(\mathbf{k}, \omega=0)$, in the metallic phase.  
Here, we approximately compute $A(\mathbf{k},\omega=0)$ as 
$A(\mathbf{k},\omega=0) \sim 
-\frac{1}{3\pi}\sum_{m=1}^{3}{\rm Im}G_{mm}(\mathbf{k}, i\omega_n \rightarrow 0)$. In Fig. \ref{fig:Ak}, we show color plots of $A(\mathbf{k},\omega=0)$. Also, shown in Fig. \ref{fig:Ak} (c) and (f) is the Fermi surface deduced from the trajectory of their peaks. In the isotropic case, $A(\mathbf{k},\omega=0)$ for $U=4.2$ has strong intensity at the Fermi surface and  the profile of the Fermi surface is consistent with that in the noninteracting case. Near the Mott transition, for $U=8.3$, the Fermi surface keeps the same shape as in the noninteracting case, although the quasiparticle peak in $A(\mathbf{k},\omega)$ is somewhat obscured at finite temperatures. In contrast, for $t' = 0.8$, the geometry of the Fermi surface drastically changes near the Mott transition. In the weak coupling regime, for $U=3.65$, the Fermi surface has a shape similar to that in the noninteracting case. As $U$ increases, for $U=7.2$, the shape becomes elongated and quite different from that in the noninteracting case. It is known that the shape of the noninteracting Fermi surface is elongated along the $k_x=k_y$ direction as $t'$ decreases \cite{imai03}. Therefore, we find that the band structure of quasiparticles becomes much more anisotropic due to the enhancement of hopping anisotropy, which is induced  by the relaxation of frustration.

\begin{figure}[bt]
\begin{center}
\includegraphics[clip,width=\figwidth]{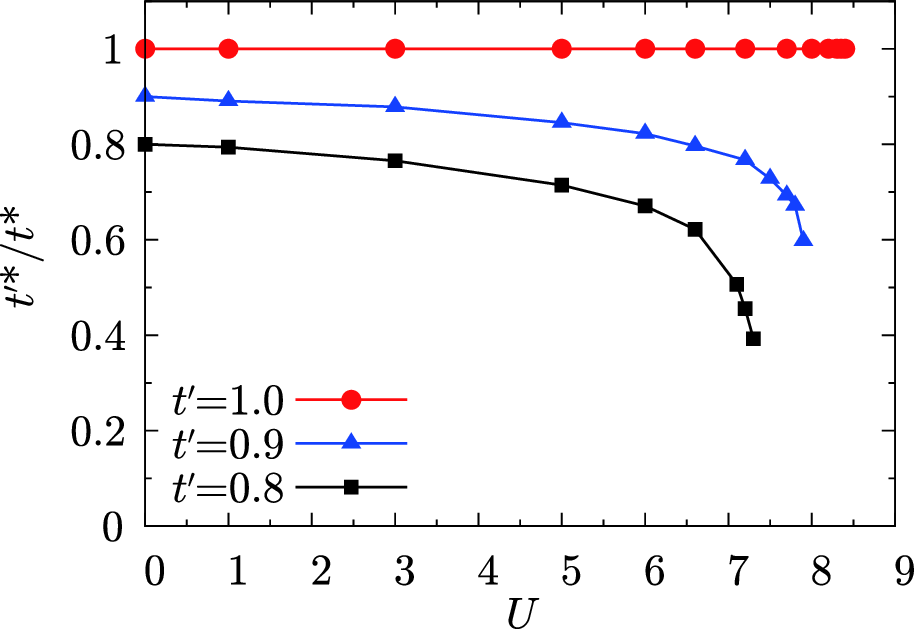}
\end{center}
\vspace{-6mm}
\caption{Ratio of the effective hopping integrals $t'^*$ and $t^*$ as a 
function of $U$ in the metallic phase. 
}
\label{fig:renhop}
\end{figure}

In order to confirm the above scenario, we investigate the renormalization of hopping integrals. In the low-energy limit, the Green's function in the effective cluster model may be given by
\begin{align}
\hat{G}(\omega)=\hat{Z}^{1/2}\left[
(\omega + \mu) \hat{1} - 
\hat{t}_\mathrm{loc}^* -
\hat{\Delta}^*(\omega) 
\right]^{-1}\hat{Z}^{1/2},
\end{align}
where $\hat{t}_\mathrm{loc}^*$ is the matrix representation of  the renormalized hopping integrals defined by 
$\hat{t}_\mathrm{loc}^*=\hat{Z}^{1/2}
(\hat{t}_\mathrm{loc} - \rm{Re}\hat{\Sigma}(\omega = 0))
\hat{Z}^{1/2}$, where 
$\hat{Z} = (\hat{1} - \partial \rm{Re}\hat{\Sigma}(\omega)
/\partial \omega |_{\omega = 0})^{-1}$. In $\hat{t}_\mathrm{loc}^*$, 
the original hopping integral $t$ $(t')$ 
is renormalized as, 
$t \to t^*$ $(t' \to t'^*)$. 
The renormalized hybridization function is defined similarly:
$\hat{\Delta}^*(\omega)=
\hat{Z}^{1/2}\hat{\Delta}(\omega)\hat{Z}^{1/2}$.

In Fig. \ref{fig:renhop}, we show the $U$-dependence of the ratio of the renormalized hopping integrals $t'^*/t^*$. 
In the isotropic case, $t'^*/t^*$ is always unity due to the symmetry requirement, giving rise to strong frustration near the Mott transition.  For the anisotropic cases of $t'/t = 0.9$ and $0.8$, $t'^*/t^*$ does not show any drastic change in the small $U$ region. However, in the vicinity of the Mott transition,  $t'^*/t^*$  rapidly decreases, implying that the anisotropy in hopping integrals is considerably enhanced there. As a result, geometrical frustration becomes weak. In other words, strong frustration developed in the nearly isotropic models triggers the strong renormalization of anisotropic hopping in the vicinity of the Mott transition.  The enhanced anisotropy deforms the band structures and develops the spin correlations. We thus conclude that the dramatic changes found around the Mott transition are due to the enhancement of anisotropy associated with the relaxation of frustration.

We note that the above mechanism for relaxation of frustration may be generic for frustrated electron systems in the vicinity of the Mott transition. For example, the reentrant Mott transition found for the anisotropic triangular lattice compound \cite{kagawa04} may be caused by the above mechanism. Namely, if the lattice anisotropy is weak, a metallic state near the Mott transition keeps the isotropic nature approximately in the weak coupling regime (high temperature) and thus behaves as a fully-frustrated model. However, as the temperature is lowered, the system enters the strong-coupling regime, where the anisotropy is enhanced and the AFM correlations are developed. This may trigger the reentrant Mott transition where the low-temperature Mott phase is accompanied by enhanced AFM correlations. 

In summary, we have studied the Mott transition in the Hubbard model on the anisotropic kagom\'e lattice by means of the CDMFT combined with CT-QMC. For anisotropic lattice systems, we have found that the quasiparticle and magnetic properties drastically change around the Mott transition; the spin correlations get strongly enhanced and the quasiparticle bands are deformed. It has been elucidated that these behaviors are due to the relaxation of frustration caused by the enhancement of anisotropy around the Mott transition. 

The authors thank Ansgar Liebsch for valuable discussions.  A part of numerical computations was done at the Supercomputer Center at ISSP, University of Tokyo and also at YITP, Kyoto University. This work was supported by KAKENHI (Nos. 21740232, No. 20104010),  the Next Generation Super Computing Project ``Nanoscience Program'' from the MEXT of Japan, the Grant-in-Aid for the Global COE Programs ``The Next Generation of Physics, Spun from Universality and Emergence'' from MEXT of Japan, and the Funding Program for World-Leading Innovative R$\&$D on Science and Technology (FIRST Program)D


\end{document}